# A Fast Template-based Approach to Automatically Identify Primary Text Content of a Web Page

Dat Quoc Nguyen, Dai Quoc Nguyen, Son Bao Pham, The Duy Bui
Human Machine Interaction Laboratory,
College of Technology, Vietnam National University, Hanoi

*Abstract*—Search engines have become an indispensable tool for browsing information on the Internet. The user, however, is often annoyed by redundant results from irrelevant web pages. One reason is because search engines also look at non-informative blocks of web pages such as advertisement, navigation links, etc. In this paper, we propose a fast algorithm called FastContentExtractor to automatically detect main content blocks in a web page by improving the ContentExtractor algorithm. By automatically identifying and storing templates representing the structure of content blocks in a website, content blocks of a new web page from the website can be extracted quickly. The hierarchical order of the output blocks is also maintained which guarantees that the extracted content blocks are in the same order as the original ones.

## I. INTRODUCTION

Nowadays, search engines have become an indispensable tool for browsing information on the Internet. While there are many useful search engines available, the users are still annoyed by redundant results from irrelevant web pages. One of the reasons is because web pages often contain non-informative blocks such as advertisements, links, etc. A search engine, which lacks effective content block detection capacity, often searches in non-informative blocks and therefore produces redundant results.

A block in a web page is often defined as a part of web page surrounded by an open tag such as <TABLE>, <TR>, <HR>, <UL>, <P>, <SPAN>, <DIV> and a matching close tag [11]. Detecting which blocks are primary text content blocks not only induces huge efficiency in storage for a search engine but also improves search efficiency in order to increase users' satisfaction. Manually marking content blocks is not a feasible solution for a search engine. In this paper, we consider the task of automatically detecting content blocks in a web page.

Web pages on the same website usually have similar structures. Furthermore, non-content blocks often situate in fixed positions. Utilizing those observations, content blocks in a web page can be automatically detected. At present, several methods have been proposed to tackle this problem including Content Extractor by Debnath et al. [11][12], noise elimination method by Yi et al. [9], InfoDiscoverer by Lin and Ho [13]. Among them, ContentExtractor appears to be the most effective algorithm to extract primary content blocks. For a web page, ContentExtractor finds content blocks by comparing each of its blocks with all blocks of web pages from the same website. The main disadvantage of this algorithm is that it is quite slow when the number of input web pages is large. Moreover, because ContentExtractor algorithm does not preserve the hierarchical order of output blocks, the extracted content blocks may not appear in the same order as the original ones. This might prevent the search engine from searching correctly an exact phrase when the phrase spans across two consecutive blocks.

In this paper, we propose FastContentExtractor - a fast algorithm to automatically detect content blocks in web pages by improving ContentExtractor. Instead of storing all input web pages of a website, we automatically create a template to store information of content blocks and possible wrongly detected blocks for later retrieval. Each block in a web page can be identified, although not always uniquely, by a traversal path in a hierarchical tree of blocks which represents the web page. A template contains a set of absolute paths of content blocks and non-content blocks having the same paths as that of content blocks. By storing the absolute paths, the hierarchical order of the output blocks is maintained which guarantees that the extracted content blocks are in the same order as the original ones. After the template for a website is stored, each newly crawled web page is compared with the template to find its primary content blocks. The number of extracted blocks and comparisons in FastContentExtractor is much smaller than that in ContentExtractor, which makes FastContentExtractor faster than ContentExtractor.

The rest of the paper is organized as follows. We summarize related materials and methods in Section II. In Section III, we described our approach. Some experiments are presented in Section IV in order to show the performance of our approach.

## II. RELATED WORKS

Several methods have been proposed to detect content blocks or non-content blocks in web pages automatically. Yi et al. [9] have proposed a tree structure which is called Site Style Tree (SST) for each website based on observations. SST is formed from the DOM tree of different web pages from the same website. Yi et al. also presented formulas for calculating the importance of each node in SST, which helps to eliminate noisy information and gives primary content. The problem of this approach appears when the number of input web pages is large. Storing million sites in the World Wide Web with SST then becomes a big issue. Kolcz and Yih [2] presented a method to identify template blocks or noisy blocks such as advertisement and navigation bars. By visually separating web pages into blocks based on vertical and horizontal lines, they calculated the block frequency for each block. If the block frequency value of a block is high, it is a template block, which is then labeled for building template model. Mehta and Madaan [10] presented an approach using regex-based template. By segmenting web pages based on the template, they could detect important sections. Vieira et al. [8] used tree mapping together with the RTDM-TD Algorithm and the Retrieve Template Algorithm for detecting the template. Lin and Ho [13] introduced a method to identify content blocks by partitioning a web page into blocks based on the <TABLE> tag. Entropy values of the terms appearing in each block are calculated and used for determining content blocks.

ContentExtractor [11][12] appears to be the most effective algorithm to identify primary informative content blocks. The input of this algorithm is a set of web pages that are assumed to have similar structure. First, the algorithm partitions each page into atomic blocks. An atomic block is a block that does not contain any block. In the next step, with an atomic block **B,** the algorithm calculates the number of web pages that contain a block similar to **B**. If block **B** occurs many times over different web pages, block **B** is considered as a non-content block, and it is removed. Otherwise, block **B** is considered a primary content block.

Figure 1 shows a block with corresponding <P> tag of a web page. This block contains four atomic sub-blocks with corresponding <span> tag (see the source code in Figure 2). ContentExtractor then partitions the block into five blocks (see Figure 3) which are the four sub-blocks and the original block with sub-blocks removed.

To identify two similar blocks, ContentExtractor uses a function to measure the similarity between two blocks. The result of this function is the cosine between two feature vectors that represent the two corresponding blocks. The feature vector of a block may include the number of images, the number of java scripts, hyperlinks and terms that appear in the block. If the returned value of the *measure function* is greater than 0.9, two blocks is considered similar. To calculate the number of web pages which contain any blocks similar to **B**, ContentExtractor algorithm compares **B** with all blocks in all input web pages.

One main disadvantage of ContentExtractor is its low speed when the number of input web pages is high. The second disadvantage is that ContentExtractor does not preserve the order of extracted content blocks. It is because the process of partitioning each webpage into atomic blocks changes the order of these blocks. Figure 3 shows the extracted blocks from the paragraph in Figure 1, which are not in the original order. This prevents an exact phrase search to be carried out properly. For example, the phrase *"the US House of Representatives"* will not be found in the extracted text.

On Sept. 27, the US House of Representatives unanimously passed a resolution recognizing The Christian Science Monitor on its centennial. The measure was sponsored by Rep. Lamar Smith (R) of Texas who once served on the Monitor staff. It was cosponsored by 40 other members of Congress.

Figure 1.  A block with <p> tag.

```
<p> On Sept. 27, the US <span
class="yshortcuts"
id="lw_1223369478_0">House of
Representatives</span> unanimously passed
a resolution recognizing <span
class="yshortcuts"
id="lw_1223369478_1">The Christian
Science Monitor</span> on its centennial.
The measure was sponsored by <span
class="yshortcuts"
id="lw_1223369478_2">Rep. Lamar
Smith</span> (R) of Texas who once served
on the Monitor staff. It was cosponsored
by 40 other <span class="yshortcuts"
id="lw_1223369478_3">members of
Congress</span>. </p>
```

Figure 2.  The source code of the block in Figure 1.

```
<span class="yshortcuts"
id="lw_1223369478_0">House of
Representatives</span>

<span class="yshortcuts"
id="lw_1223369478_1">The Christian
Science Monitor</span>

<span class="yshortcuts"
id="lw_1223369478_2">Rep. Lamar
Smith</span>

<span class="yshortcuts"
id="lw_1223369478_3">members of
Congress</span>

<p> On Sept. 27, the US unanimously
passed a resolution recognizing on its
centennial. The measure was sponsored by
(R) of Texas who once served on the
Monitor staff. It was cosponsored by 40
other . </p>
```

Figure 3.  Extracted sub-blocks from the paragraph in Figure 1.

III.  OUR APPROACH

In this section, we describe our FastContentExtractor algorithm that extends ContentExtractor algorithm. By building and storing a template for each website, we can later extract the primary content of any web page from that website.

Different from ContentExtractor, our FastContentExtractor contains two phases: the preparation phase and the detection phase. At the preparation phase, FastContentExtractor collects a set of web pages from a given website to automatically generate a template to describe content blocks (see Figure 4). This phase is carried out infrequently. Similar to ContentExtractor algorithm, first, we identify content blocks from atomic blocks of the web pages. We then store the traversal path of these blocks along the hierarchical trees of blocks representing the web pages. The traversal path of a block is a string of the form "$tag_1.tag_2.tag_3....tag_n$" where the block with corresponding

$tag_{i+1}$ is a sub-block of the block with corresponding $tag_i$, $tag_n$ is the tag of an atomic block, and $tag_1$ is the most generic tag "HTML". For example, "HTML.BODY.TABLE.TR.P" is the string representing the traversal path to a block. The advantage of this way to describe a block is the independence of its position in the web page. The disadvantage of this way is that it does not provide a unique way to identify a block in a web page. Thus, two different blocks may have the same traversal path. For this reason, we also store in the template the content of non-content blocks which have the path as content blocks in order to correctly identify content blocks in a new web page later.

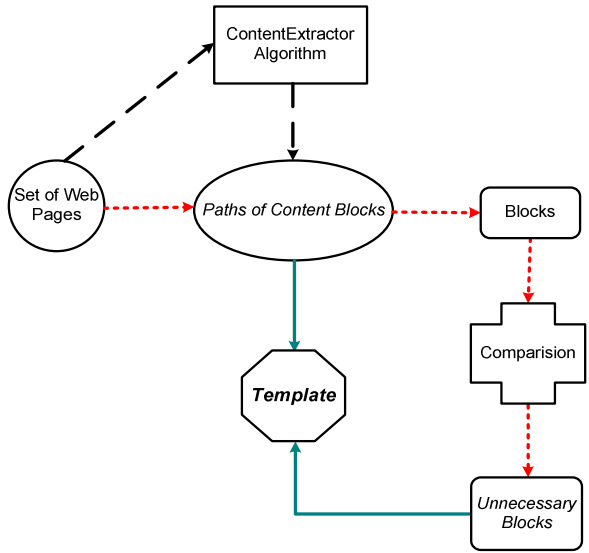

Figure 4. The preparation phase of the FastContentExtractor algorithm.

In the detection phase, by using the stored template of the corresponding website, content blocks of a new web page can be detected quickly (see Figure 5). Only blocks of the new web page having the same paths as the paths stored in the template are extracted. Denoting **P** as the set of paths storing in the template and **B** is a block with a path in **P**, the extraction rules are as follows:

*if* the path of all sub-blocks (if any) of **B** are in **P, then** the whole block **B** is extracted;

*if* **B** contains a block **B'** with a path not in **P, then**
*if* the path of all sub-blocks (if any) of **B'** are not in **P, then** block **B** is extracted without **B'**;
*otherwise* block **B** is extracted including **B'**.

An extracted block is not necessarily an atomic block. Each extracted block is then compared with non-content blocks stored in the template. If the block is considered similar to a non-content block, it is considered as non-content block. Otherwise, it is considered as content block and its text is extracted as the primary text content of the web page. For example, in Figure 2, all of blocks with corresponding <p> tag and sub-blocks with corresponding <span> tag are considered as content blocked and are extracted.

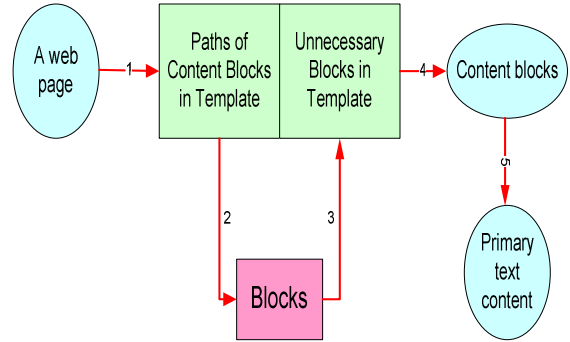

Figure 5. The detection phase of the FastContentExtractor algorithm.

It can be seen that the number of comparisons in FastContentExtractor is much smaller than that in ContentExtractor. Moreover, while ContenExtractor algorithm does not keep primitive structure of blocks in its output, by using the paths of content blocks, FastContentExtractor retains primitive structure of blocks to keep information content intact.

## IV. RESULTS

We compare the execution time and accuracy between our FastContentExtractor algorithm (FastCE) and our own implementation of ContentExtractor algorithm (CE). Both FastCE and CE take a set of web pages from the same site as input and output the corresponding text content or primary content blocks. In this experiment, we use both Vietnamese and English websites as shown in Table I.

TABLE I. THE WEBSITES USED IN THE EXPERIMENT AND THEIR CORRESPONDING NUMBER OF WEB PAGES

| Address | Number of web pages |
|---|---|
| dantri.com.vn | 337 |
| kenh14.vn | 269 |
| thanhnien.com.vn | 290 |
| vietnamnet.vn | 365 |
| news.yahoo.com | 115 |
| cnn.com | 191 |
| news.bbc.co.uk | 106 |
| nytimes.com | 100 |

We use between 20 to 30 web pages at the preparation phase to generate the template for each website.

### A. Execution time

In order to compare the execution time between FastCE and CE, we define the following terms:
- *NumBlockTemp* refers to the number of blocks that are used to compare to decide whether a block is a content block.
- *NumBlock* refers to the averaged number of blocks that are generated by each approach for each webpage in input data set. *NumBlock* is the number of atomic blocks for CE, and is the number of blocks that are generated by using the paths of content blocks for FastCE.

- *PerTime* refers to the averaged execution time for each web page from the input data set. *PerTime* includes the time taken to extract blocks and to compare the extracted blocks with the blocks stored in the template.

Because the number of blocks in the template and the number of extracted blocks in FastCE approach is smaller compared to that in CE, the comparison time between blocks is smaller for FastCE approach. Similarly, the amount of time taken to extract blocks in FastCE approach is smaller than that in CE. Therefore, the overall execution time in FastCE approach is smaller compared to CE approach as illustrated in Table II and Figure 6. In fact the runtime for FastCE is significantly better compared to that of CE across all websites experimented.

TABLE II. EXECUTION TIME OF CE AND FASTCE

| Address | NBT / NB / PerTime in CE | NBT / NB / PerTime in FastCE | Improvement on execution time |
|---|---|---|---|
| dantri.com.vn | 86 / 319 / 1.914 | 14 / 41 / 0.964 | 198.55% |
| kenh14.vn | 247 / 500 / 18.4 | 26 / 46 / 1.39 | 1323.7% |
| thanhnien.com.vn | 111 / 326 / 1.817 | 19 / 14 / 0.703 | 258.5% |
| vietnamnet.vn | 23 / 121 / 0.563 | 3 / 22 / 0.527 | 106.8% |
| news.yahoo.com | 114 / 171 / 1.883 | 34 / 48 / 0.938 | 200.7% |
| cnn.com | 112 / 266 / 2.924 | 20 / 15 / 2.002 | 146.5% |
| news.bbc.co.uk | 77 / 174 / 1.401 | 16 / 45 / 0.565 | 247.9% |
| nytimes.com | 318 / 146 / 2.273 | 58 / 17 / 1.557 | 146% |

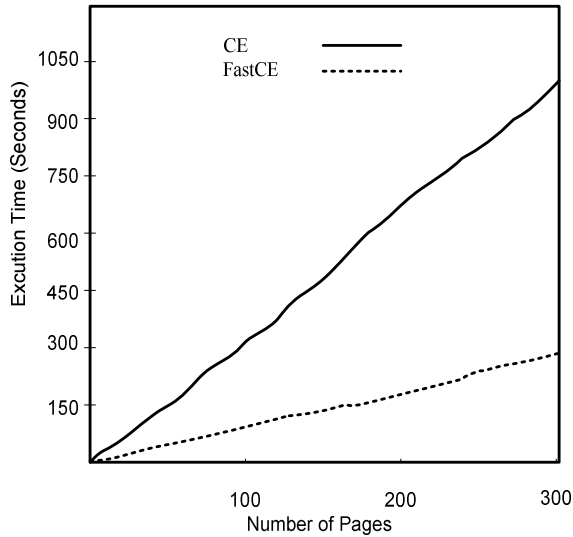

Figure 6. Average Processing Time for CE and FastCE.

B. *Accuracy*

*1) Block level accuracy.* Similar to Debnath et al. [11], we use $B_{Fmeasure}$ as a metric to compare the accuracy:

$$B_{F_{measure}} = \frac{2 * B_{recall} * B_{precision}}{B_{recall} + B_{precision}}$$

$B_{recall}$ is defined as the ratio between the number of content blocks extracted and the actual number of content blocks while $B_{precision}$ is defined as the ratio between the number of content blocks extracted and the total number of extracted blocks.

TABLE III. $B_{FMEASURE}$ FOR CE AND FASTCE ON A NUMBER OF WEBSITES

| Address | $B_{Fmeasure}$ in CE | $B_{Fmeasure}$ in FastCE |
|---|---|---|
| dantri.com.vn | 0.97 | 0.97 |
| kenh14.vn | 1.00 | 1.00 |
| thanhnien.com.vn | 0.90 | 0.89 |
| vietnamnet.vn | 0.83 | 1.00 |
| news.yahoo.com | 0.86 | 0.92 |
| cnn.com | 0.91 | 0.75 |
| news.bbc.co.uk | 0.88 | 0.94 |
| nytimes.com | 0.90 | 0.91 |

Table III shows the measure of block level accuracy for CE and FastCE on a number of websites. As can be seen from the table, the accuracy of FastCE is similar to that of CE.

*2) Word level accuracy.* In this section, we execute the comparison based on word levels. We use $W_{Fmeasure}$ as a metric to compare the accuracy between FastCE and the CE:

$$W_{F_{measure}} = \frac{2 * W_{recall} * W_{precision}}{W_{recall} + W_{precision}}$$

$W_{recall}$ is defined as the ratio between the number of words in extracted primary content and number of words in original primary content. $W_{precision}$ is defined as the ratio between the number of words in extracted primary content and total number of extracted words.

TABLE IV. $W_{FMEASURE}$ FOR CE AND FASTCE ON A NUMBER OF WEBSITES

| Address | $W_{Fmeasure}$ in CE | $W_{Fmeasure}$ in FastCE |
|---|---|---|
| dantri.com.vn | 0.978 | 0.991 |
| kenh14.vn | 1.00 | 1.00 |
| thanhnien.com.vn | 0.978 | 0.992 |
| vietnamnet.vn | 0.99 | 0.999 |
| news.yahoo.com | 0.89 | 0.958 |
| cnn.com | 0.99 | 0.99 |
| news.bbc.co.uk | 0.957 | 0.96 |
| nytimes.com | 0.966 | 0.966 |

It can be seen from Table IV that FastCE performs as accurately as CE for most of the websites experimented.

## V. CONCLUSION

We proposed in this paper FastContentExtractor - a fast approach for extracting primary content of web pages. FastContentExtractor extends ContentExtractor algorithm by building templates for each website at hand where the template contains paths to the content blocks as well as distinct non-content blocks. Experiments on both Vietnamese and English websites have demonstrated the advantage of FastContentExtractor over ContentExtractor. In particular, FastContentExtractor outperformed ContentExtractor by a high margin in runtime while maintaining the accuracy. In addition, FastContentExtractor keeps text information content intact which allows the exact phrase search to perform correctly.

ACKNOWLEDGEMENT

This work is partly supported by the research project No. QC.08.17 granted by Vietnam National University, Hanoi.